\documentclass[9pt,twocolumn,twoside]{osajnl}

\journal{ol} 
\usepackage[]{siunitx}
\newcommand{\E}{\text{e}}
\newcommand{\I}{\text{i}}
\newcommand{\refeq}[1]{Eq.~(\ref{#1})}
\newcommand{\reffig}[1]{Fig.~\ref{#1}}

\setboolean{shortarticle}{true}

\title{Shaping convex edges in borosilicate glass by single pass perforation with an Airy beam}
\author[1,2*]{David Sohr}
\author[2]{Jens Ulrich Thomas}
\author[1]{Stefan Skupin}

\affil[1]{Institut Lumière Matière, UMR 5306 Université Lyon 1 - CNRS, Université de Lyon, 69622 Villeurbanne, France}
\affil[2]{SCHOTT AG, Hattenbergstrasse 10, 55120 Mainz, Germany}

\affil[*]{Corresponding author: david.sohr@univ-lyon1.fr}

\begin{abstract}
We demonstrate curved modifications with lengths of up to \SI{2}{mm} within borosilicate glass produced by single 1030 nm picosecond laser shots with an Airy beam profile. Plasma ignition in the side lobes of the beam as well as surface damage prove to be the crucial limitations for confined bulk energy deposition on a curved trajectory. A combined experimental and numerical analysis reveals optimum laser parameters for confined bulk energy deposition. This way we achieved single pass cutting of a \SI{525}{\micro m} thick glass sheet with a well defined convex edge down to a bending radius of \SI{774}{\micro m}.
\end{abstract}

\setboolean{displaycopyright}{true}

\begin{document}

\maketitle

For reasons reaching from practical purposes, such as improved handling and durability, to decorative considerations, processed glass is often required to have a seamed or round edge. Currently this requires an extra grinding step to obtain for example the most common curved shape, the so called ``c-cut''.
For the preceding processing step of cutting the glass to shape, ultra short pulse lasers used with beam profiles with an elongated focal volume (line focus) are increasingly replacing conventional cutting tools~\cite{Mishchik.2017, Bergner.2018, Jenne.2020d}. These straight line foci can be used to modify the workpiece throughout its entire depth with one single laser shot, effectively perforating it instantaneously, instead of having to dig into the material, as for example by ablation~\cite{Mathis.2012}. Avoiding the need to remove material makes this perforation a very clean process. The most prominent beam shape in this context has been the Bessel-Gauss beam~\cite{Marcinkevivcius.2001, Bhuyan.2010}. With its non diffracting, self healing character it provides a highly uniform and robust line focus within the material. Deformation of the laser beam shape through nonlinear propagation effects can be minimized by using high NA focusing optics, feeding energy to the line focus increasingly from the side, and by employing bursts of laser pulses~\cite{Mishchik.2017}. The laser energy deposited in a narrow channel acts to create permanent modifications reaching from structural changes to voids and cracks, in any case weakened zones compared to the pristine material. By laterally joining several of these weak zones, e.g. by mechanical cleaving or etching, the glass can then be separated. This way single pass laser glass cutting up to a thickness of \SI{12}{mm} has been demonstrated~\cite{Jenne.2020d} and this thickness was fundamentally limited only by the laser pulse energy and the size of the optics. The results of such a cutting process are straight, generally vertical edges. Even though the cleaved edges with an root-mean-square roughness of one micron resemble a fine polish, the sharp 90 degree edge has usually to be removed in a subsequent edge shaping step to avoid for example the danger of cut injuries.

In this Letter we report on combining cutting and edge shaping of glass in one laser process. To this end, we replace the Bessel-Gauss beam in the aforementioned micromachining setup with an Airy-Gauss beam to directly achieve a curved edge as the result of the cutting process.
The Airy beam is the most prominent non diffracting beam following a curved trajectory, that is to say its main lobe appears to be accelerated in the transverse plane during propagation. After its theoretical description in 1979~\cite{Berry.1979}, the optical, finite-energy Airy-Gauss beam featuring a curved line focus was first demonstrated experimentally in 2007~\cite{Siviloglou.2007}. While the applications that followed reach from 1D light sheets~\cite{Yang.2014} to light bullets formed by a ring-Airy~\cite{Panagiotopoulos.2013}, the most commonly treated case is still the 2D case given by the linear combination of cubic phases along two perpendicular axes in the far field~\cite{Mathis.2012, Froehly.2011, Polynkin.2009}.
As the first of a whole group of accelerating beams~\cite{Efremidis.2019}, the 2D Airy beam has been used before to create long, curved plasma channels (ca.\ \SI{1}{m} length) in air~\cite{Polynkin.2009}.

To our knowledge there has not yet been a case of the Airy beam being used for creating extended mechanical damage zones in dense media, analogous to the Bessel beam perforation in dielectrics.
The demonstration of shaping convex edges with the Airy beam has been limited to ablation of diamond and silicon for a thickness of up to ca.\ \SI{100}{\micro m}~\cite{Mathis.2012}, requiring many passes for a single convex profile and producing significant pollution due to the material removed during the laser process. The hard focusing (NA~0.8) in this previous study provided a strong curvature with bending radius $r = \SI{120}{\micro m}$, but also limited the length of the line focus. Curved refractive index modifications produced by single shot Airy beams in glass were demonstrated up to a length of only up to ca.\ \SI{20}{\micro m}~\cite{Froehly.2011}.
Here, we first demonstrate curved bulk modifications in borosilicate glass up to lengths in the mm range. Second, by means of experiments and numerical simulations, we investigate the influence of the laser parameters, in particular pulse duration and number of pulses in a burst, in order to optimize the bulk energy deposition. Finally, we confirm that the optimized laser process causes sufficient mechanical damage throughout the glass sheet to allow subsequent separation, and report the creation of a well defined convex edge.

\begin{figure}
\centering
\includegraphics[width=\linewidth]{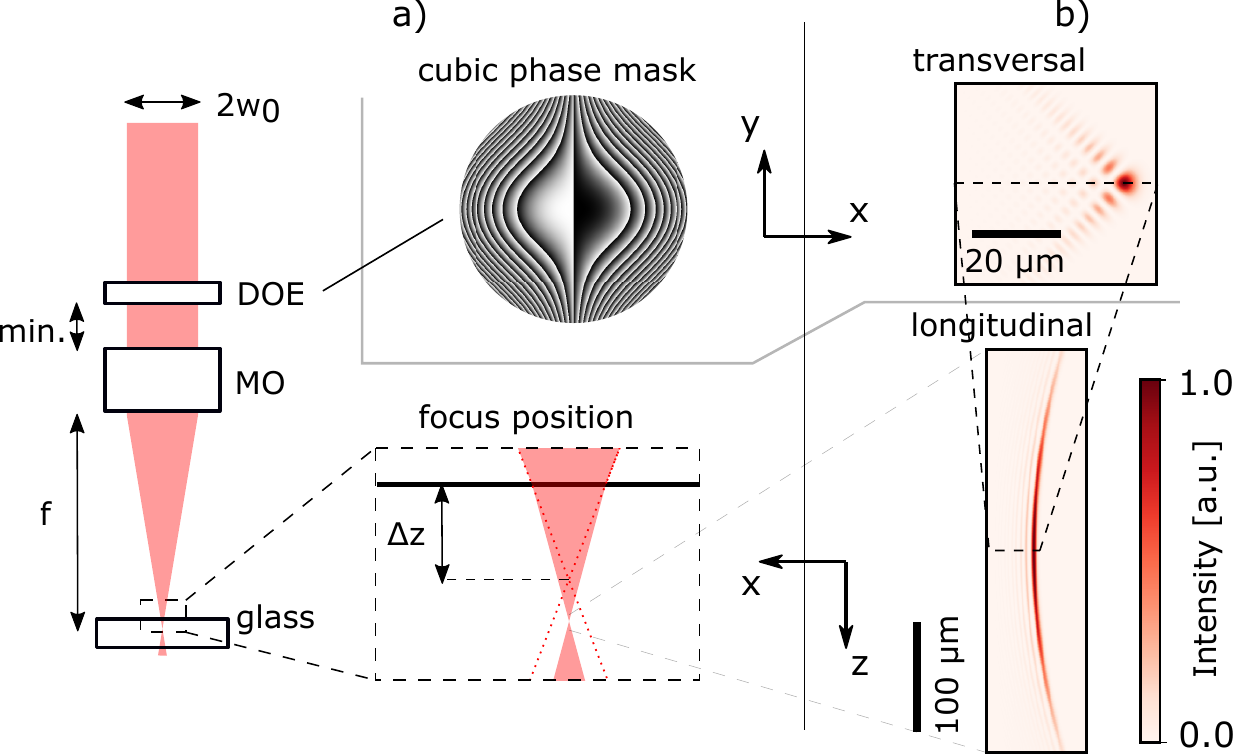}
\caption{Experimental 2f-setup (a) used for creating an Airy-Gauss beam (see text for details) and intensity profiles (b) according to \refeq{eq:fullAiry} for $f = \SI{10}{mm}$. Throughout the paper, the focus position is given as the shift $\Delta z$ of the glass surface with respect to the focus position in air (dotted line).}
\label{fig:2fsetup}
\end{figure}

We begin with the paraxial description of the Airy-Gauss beam. For a Gaussian laser beam with an optical wavelength $\lambda$ and a half beam width (at $1/\E^2$ intensity) $w_0$, on which a cubic phase $\exp[\I\beta^3(x^3/3+xy^2)/\sqrt{2}]$ with a scale factor $\beta$ is added and which is then focused by an optical element with an effective focal length $f$, the propagation depends on the transverse length scale factor $x_0 =\sqrt{2} f \beta/k$, the dimensionless longitudinal coordinate $\zeta = 2z/(k x_0^2)$ and the confinement factor $a = 1/(w_0^2 \beta^2)$~\cite{Polynkin.2009}.
Here, $k=2\pi n/\lambda$ with $n$ being the refractive index of the medium.
The resulting line focus given by the main lobe of the complex electric field envelope~\cite{Siviloglou.2007}
\begin{align}
      E(\zeta,x,y) & =
			\mathrm{Ai}\!\left(\frac{x-y}{x_0} - \frac{\zeta^2}{4} + \I a \zeta\right)
			\mathrm{Ai}\!\left(\frac{x+y}{x_0} - \frac{\zeta^2}{4} + \I a\zeta\right) \nonumber \\
			& \quad  \times \exp \! \left[a \left(\frac{2x}{x_0} - \xi^2 \right)
			- \I \xi  \left(\frac{\xi^2}{6} + a^2 + \frac{x}{x_0}\right)\right]\,,\label{eq:fullAiry}
\end{align}
follows a parabolic trajectory in the $xz$ plane with a quadratic coefficient of $q = 1/(k^2x_0^3)$, see \reffig{fig:2fsetup}\,b). At the vertex of the parabola this corresponds to an effective radius $r = k^2 x_0^3/2$ and the angle $\alpha$ between the optical axis and the line focus varies along $z$ as ${\alpha(z) = \arctan(2q z)}$. Over a length of $l \approx 2 w_0 f^2 \beta^3/k$ the intensity of the main lobe exceeds $1/\E^2$ of its maximum.

\begin{figure}[t]
\centering
{\includegraphics[width=\linewidth]{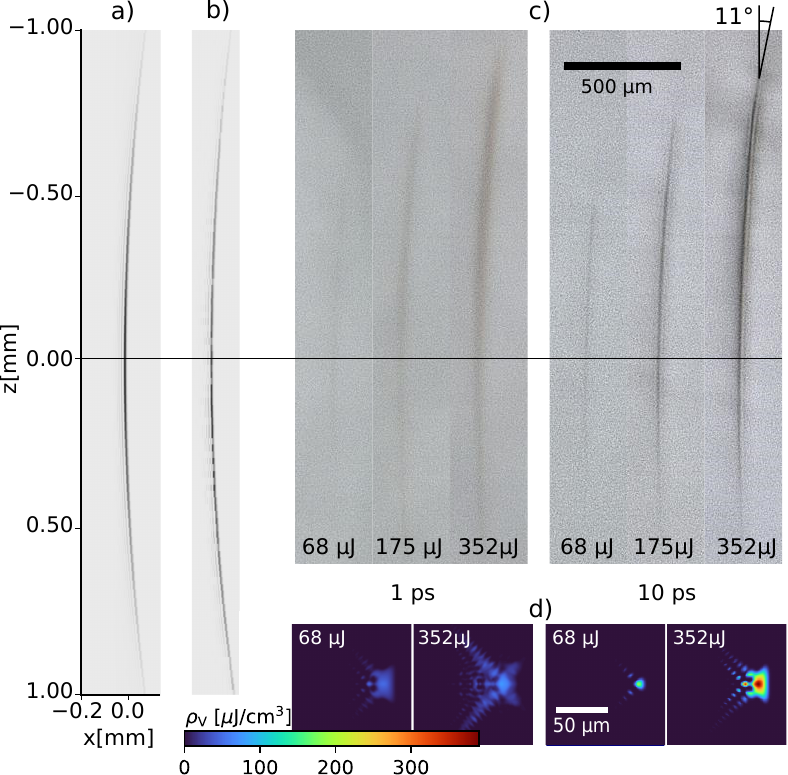}}
\caption{The linear propagation of an Airy-Gauss beam with focal length $f = \SI{20}{mm}$ in air observed experimentally (b) agrees very well with the expected profile (a) given by \refeq{eq:fullAiry}. Light microscopy cross sections of permanent in-volume modifications in glass (c) also clearly follow the same trajectory. Corresponding deposited energy densities in the focal plane (d) obtained from numerical simulations confirm the higher confinement of deposited energy for long pulses. The $z$ scale of the profiles in air in (a,b) is stretched by the refractive index of borosilicate glass $(n=1.46)$ to mimic propagation in glass. 
}
\label{fig:VolProf_f20}
\end{figure}

We used a 2f setup for creating the Airy-Gauss beam, as shown in \reffig{fig:2fsetup}\,a) and similar to~\cite{Mathis.2012}, imaging the cubic phase produced by a diffractive optical element (DOE: Airy Beam Generator, Holo/Or) with a microscope objective (MO). As the front focal plane of the MO lies within the housing of the MO, the DOE could not be placed exactly at this plane. Rather the DOE has been placed as close to the MO as the mechanical parts allowed (effective distance \SI{4}{mm}). The setup using the MO as described here resulted in the expected propagation behaviour of a Airy-Gauss beam as shown in \reffig{fig:VolProf_f20}\,a/b). In contrast placing the DOE at the (accessible) front focal plane of aspheric lenses lead to a much more distorted beam propagation.

The laser used for micromachining, an Amphos 200XHE, has a wavelength $\lambda$ of \SI{1030}{nm}, an $M^{2}$ of 1.1, a beam radius $w_0$ of \SI{2.65}{mm}, and can emit bursts of 1 to 4 pulses with a delay of \SI{25}{ns} and a pulse duration $t_{\rm p}$ between \SI{1}{ps} and \SI{10}{ps}. The scale factor of the cubic phase $\beta$ was fixed at $3^{1/3}$ mm$^{-1}$. Thus with focal lengths of \SI{10}{mm} and \SI{20}{mm} it was possible to reach an effective bending radius in air of \SI{0.70}{mm} and \SI{5.6}{mm} respectively.

Sheets of borosilicate glass (SCHOTT {Borofloat\textsuperscript{\textregistered} 33}) of various thicknesses were placed horizontally on a motorized XY stage with the laser beam coming from the top. The orientation of the Airy beam with respect to the mechanical axes and the exact position of the focus was determined by examining the ablation pattern at the glass surface. The focus of the Airy beam was subsequently placed within the glass volume by decreasing the distance between the focusing optics and the glass sheet by a defined distance $\Delta z$ (see \reffig{fig:2fsetup}). The glass was then moved horizontally with respect to the laser beam while picking single laser pulses, resulting in discrete modifications, separated along the lateral translation direction by a constant distance, a so-called pitch, of 10, 25 or \SI{50}{\micro m} respectively. This translation direction was oriented perpendicularly to the acceleration direction of the beam, thus maximising the curvature of the cross section profile. Experiments were performed for various combinations of pulse duration, burst mode, pulse energy, focus position and pitch.

In order to first examine the pure volume modification, we placed the focus deeply within a $\SI{6.5}{mm}$ thick glass sheet ($\Delta z = \SI{2}{mm}$) thus avoiding plasma ignition at the surface. We inspected the longitudinal profile of the volume modifications by cutting the glass perpendicularly to the above mentioned lateral translation direction, polishing it and looking down this translation direction using reflected light microscopy (Keyence VHX 6000). For these profile measurements we took special care to avoid double shots, implementing a pulse on demand operation.
We found that the volume modifications show the expected curvature for a wide range of experimental parameters (see \reffig{fig:VolProf_f20}). With increasing pulse energy the trajectory remains unchanged while the length and lateral extent of the modifications increase, for $f = \SI{20}{mm}$ reaching lengths of up to \SI{2}{mm} and a maximum angle $a_{max} = 11^{\circ}$. For \SI{1}{ps} pulses the modifications remain broad and indistinct. More confined and stronger modifications are observed for longer pulse durations. The experimental profiles show a shift of the maximum intensity or damage respectively with respect to the vertex of the trajectory, probably partly due to a slight misalignment of the laser beam. In particular, however, we observe an increasingly asymmetric damage distribution along the parabola with increasing pulse energy, with the most extensive damage shifted towards the laser source and a long tail away from it.

To increase our understanding of the material damage we simulated the energy deposition by using an unidirectional pulse propagation model accounting for all relevant nonlinear propagation effects, in particular the laser-generated conduction electrons~\cite{Berge:rpp:70:1633,dematteo:apl:107:181110}.
The preferential energy deposition in front of the focus observed in the experiments is in fact well captured by our simulations, and can be explained straightforwardly by a combined action of Kerr self-focusing and nonlinear absorption due to plasma generation, in particular in the side lobes of the Airy beam.
While the total energy deposition was larger for short pulses with $t_p = \SI{1}{ps}$, the energy density was much higher for the longer pulses with $t_p = \SI{10}{ps}$ for all pulse energies considered (see \reffig{fig:VolProf_f20}d). 
For the short pulses the laser affected zone is much broader due to intensity clamping~\cite{LIU2002189,AndreasSchmittSody.2016}, resulting in high absorption, but a low energy confinement. This corroborates the experimental observation of more confined and stronger modifications for the \SI{10}{ps}-pulses in \reffig{fig:VolProf_f20}c), and indicates that longer pulses are favorable for cutting applications. However, it is known that avalanche ionization is dominant compared to multiphoton ionization for long pulses, which leads to a less deterministic process~\cite{Sanner.2010}, disadvantageous for the intended process.

Moreover, in the above reasoning only bulk modifications are considered (the focus lying deeply within a thick glass sheet), and any surface modifications are excluded. The threshold for laser induced damage within the volume is supposed to be higher than that for the surface~\cite{Bloembergen:73}. Indeed, we determined the surface damage threshold intensity for Borofloat\textsuperscript{\textregistered} 33 as \SI[separate-uncertainty  = true]{1.07 \pm 0.12}{TW/cm^2} for $t_p = \SI{10}{ps}$, similar to previous values for borosilicate glass~\cite{Nieto.2015}. On the other hand, for a peak intensity of \SI{2.4}{TW/cm^2} (theoretical, for linear propagation) we hardly see any volume modification even for long pulses in \reffig{fig:VolProf_f20}c), \SI{10}{ps}, \SI{64}{\micro J}.
For the 2D Airy in our experiments the ratio between main and side lobe intensity is only about 2, compared to above 5 for a Bessel beam with similar focusing conditions~\cite{Mishchik.2017}. In previous Airy beam micromachining studies this was not an issue, because the side lobes were propagating in air~\cite{Mathis.2012}. When focusing within the glass volume, however, the balance between surface and volume damage is of particular importance due to this low focal contrast.

\begin{figure}
\centering
{\includegraphics[width=\linewidth]{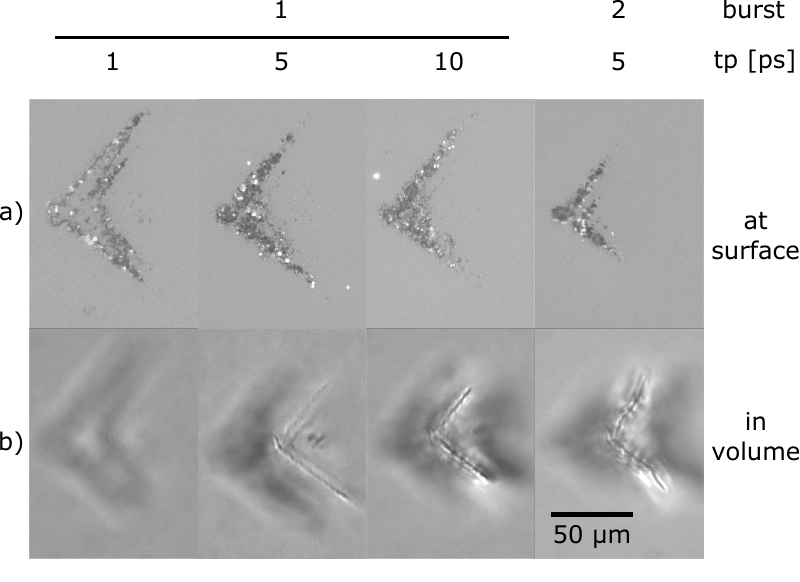}}
\caption{Light microscopy top views of permanent modifications (a) at the surface and (b) at a depth of \SI{60}{\micro m} within the volume of the glass with focal length $f = \SI{10}{mm}$, burst energy $E_{burst} = \SI{204}{\micro J}$ and focus position  $\Delta z = \SI{200}{\micro m}$ for varying pulse duration $t_p$ and burst configuration. Both reflected (a) and transmitted (b) light microscopy were used for optimal contrast, still the blurred imaged of the surface modification can be seen in (b).}
\label{fig:Trans}
\end{figure}

To study the relationship between surface and volume modifications for the Airy beam in more detail, we chose a focus position closer to the surface of the sample ($\Delta z= \SI{200}{\micro m}$) and observed the resulting modifications with both reflected and transmitted light microscopy (Zeiss Axio Imager).
By changing the peak intensity while keeping the fluence constant, we monitored surface and bulk damage for various pulse configurations, shown in \reffig{fig:Trans}. For this we also considered a burst mode operation of the laser, as this can enhance the laser energy deposition in the bulk~\cite{Mishchik.2017}. We could obtain a strong mechanical damage within the volume, even close to the surface, for $t_p = \SI{5}{ps}$ and a burst with 2 pulses.  Longer pulses on the other hand show weaker volume modifications close to the surface even though there is a significant surface damage. In fact, we suspect that the plasma ignited at the surface may even lead to a shadowing effect, suppressing volume modifications close to the surface.

We note that our pulse durations lie in the intermediate range where the scaling of the surface damage threshold changes from a more intensity dominated regime for short pulse durations to a more fluence dominated regime for long pulse durations~\cite{Stuart.1996}. It may well be that a different scaling with intensity or fluence between surface and volume damage is the main reason for the observed optimum at intermediate pulse durations and burst. 

For the separation experiments we used glass of $\SI{525}{\micro m}$ thickness, allowing perforation with $f = \SI{10}{mm}$. The glass pieces were subsequently separated, if possible, either by mechanical breaking (cleaving) or etching with potassium hydroxide~\cite{Gottmann.2013}. The resulting profile was recorded using confocal light microscopy in addition to the cross section analysis.
Although it was possible to cleave the glass along a line perforated with the Airy beam, the new surface follows only partially the perforations created by the laser process: Often the crack "cuts short" and runs trough the convex side, building up as concoidally fractured bits within the concave side of the intended cut. The process then rather resembles laser scribing \cite{Nisar.2013} rather than proper full volume cutting.

\begin{figure}
\centering
{\includegraphics[width=\linewidth]{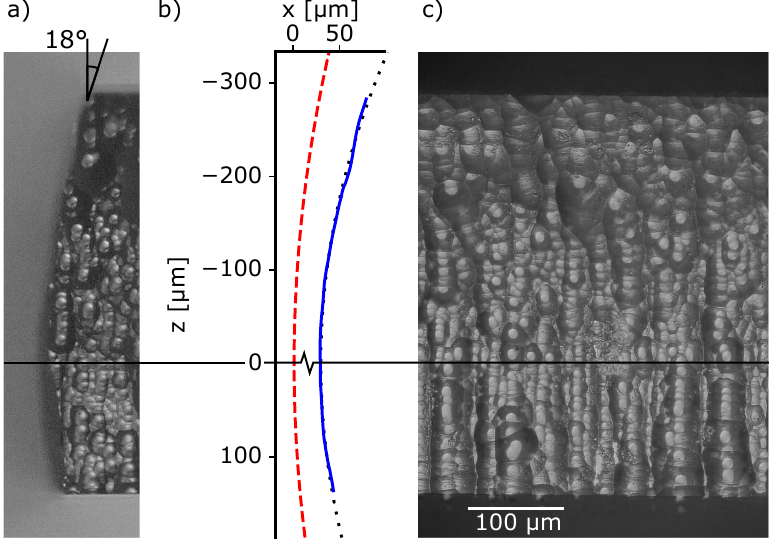}}
\caption{Result of cutting a \SI{525}{\micro m} thick glass sheet by perforation with an Airy beam and subsequent etching (\SI{50}{\micro m} etching depth). A cross section view (a) parallel to the line of laser modifications shows the curved profile of the cut. The average height profile (b) measured by confocal light microscopy (solid blue) and its parabolic fit (dotted black) exhibit a considerably stronger curvature than the Airy profile (dashed red), which can be attributed to the etching process. The side view onto the convex side (c) shows a coarser structure at the bottom, probably due to less extensive damage leading to more localized etching.}
\label{fig:AiryEtch}
\end{figure}

In contrast, \reffig{fig:AiryEtch} shows the fully expressed convex side obtained after separation by etching after laser processing with an Airy beam with $f = \SI{10}{mm}$, using a burst of two pulses with each a width of $t_p = \SI{5}{ps}$ and an energy of $E_{burst} = \SI{228}{\micro J}$, perforating with a pitch of \SI{10}{\micro m}. The convex profile follows a parabola with an effective bending radius $r_{eff} = \SI{774}{\micro m}$, which is smaller than the radius of the Airy main lobe trajectory ($r = \SI{1.5}{mm}$). This stronger curvature and the large top angle $a_{max} = 18^{\circ}$ can be understood as an extra taper angle resulting from the etching process adding to the curved Airy profile.
Due to the preferential absorption in the upper part of the glass sheet (as discussed above) we had to choose a focus position below the center of the glass sheet ($\Delta z \approx \SI{220}{\micro m}$) in order to be able to perforate the glass sheet completely. The side view in \reffig{fig:AiryEtch}c) corroborates a more extensive damage in the upper part of the glass sheet and a weaker damage in the lower part, as the edge surface below the focus shows a coarser structure. This coarser structure can be explained by a more localized etching process due to less extensive (and more variable) laser damage.

We note that the opposite side of the cut (not shown here) does not exhibit the expected concave profile. Instead, more material is removed on the upper side of the glass sheet, thus resulting in a nearly planar, inclined edge. We attribute this finding to an increased removal rate during etching in the glass volume modified by interactions with the Airy side lobes in addition to the previously mentioned taper angle.

In conclusion, we have demonstrated single pass cutting of glass sheets of \SI{525}{\micro m} thickness with an Airy-Gauss beam, resulting in a convex edge. As for other cutting processes the choice of laser parameters is crucial for the separability of the cut. However, the low focal contrast the Airy beam renders the balance between surface and volume damage particularly important. By means of a combined experimental and numerical study we identified suitable parameters for confined bulk energy deposition and moderate surface damage, in particular with respect to the side lobes of the Airy profile.
The etching process necessary for separation further reduced the achievable bending radius by almost a factor two, well beyond that of the Airy trajectory, which itself is limited by the tradeoff between focus length and curvature.
We expect that this first report of single pass curved glass cutting will trigger further research activities on micromachining with accelerating beams, aiming for thicker glass and even more pronounced curvatures.  A promising alternative candidate for producing curved modifications could be the accelerating Bessel beam~\cite{Zhao.2013, Chremmos.2013}, which distributes the side lobe intensity more evenly, thus yielding a narrower focus and a higher focal contrast, or an Airy beam with optimized intensity distribution~\cite{Barwick.2011}.

Numerical simulations were performed using HPC resources from GENCI (Grants \# A0070506129 and A0080507594).
SS acknowledges support by the Qatar National Research Fund (Grant \# NPRP 12S-0205-190047).


\textbf{Disclosures.} DS: SCHOTT AG (E,P), JUT: SCHOTT AG (E,P), SS: SCHOTT AG (F)

\bibliography{airy.bib}
\end{document}